\documentclass[pra, reprint, twocolumn, superscriptaddress, amsmath, amssymb, aps, ,floatfix]{revtex4-2}

\usepackage[english]{babel}
\usepackage{graphicx}
\graphicspath{{./Images/},{./ImagesAppendix/},{./images/},{./imagesAppendix/}}

\usepackage{amsmath, amssymb, amsfonts}
\usepackage{physics}
\usepackage{bm}
\usepackage{braket}
\usepackage{array}
\usepackage{multirow}
\usepackage{float}
\usepackage{commath}
\usepackage{verbatim}
\usepackage{color}
\usepackage{enumitem}
\usepackage{array}   
\usepackage{dsfont}
\usepackage{mathtools}
\usepackage{soul}
\usepackage[colorlinks=true, citecolor=darkBlue, linkcolor=darkBlue, urlcolor=blue]{hyperref}
\usepackage{amssymb}
\usepackage{tabularx}
\usepackage{booktabs}
\definecolor{darkBlue}{rgb}
{0.08, 0.13, 0.4}

\usepackage{cancel}

\usepackage{tikz}
\usetikzlibrary{positioning, arrows.meta}
\usepackage{pgfplots}

\definecolor{THc}{rgb}{0.9,0.3,0.2}

\newcommand{\canc}[1]{}

\begin{document}

\title{Magic transfer in quantum spin chains}
\author{Antonio Palamara}
\thanks{These authors contributed equally to this work}
\affiliation{Dipartimento di Fisica, Universit\`a della Calabria, 87036 Arcavacata di Rende (CS), Italy}
\affiliation{INFN, Sezione LNF, gruppo collegato di Cosenza}

\author{Chad Nelmes}
\thanks{These authors contributed equally to this work}
\affiliation{School of Physics, Engineering and Technology, University of York, York, YO10 5DD, United Kingdom}
\affiliation{York Centre for Quantum Technologies, University of York, York, YO10 5DD, United Kingdom}

\author{Enrico C. Domanti}
\affiliation{Instituto de Física Teórica, UAM-CSIC, Universidad Autonoma de Madrid, Cantoblanco, 28049 Madrid, Spain}

\author{Francesco Perciavalle}
\affiliation{Dipartimento di Fisica, Universit\`a della Calabria, 87036 Arcavacata di Rende (CS), Italy}
\affiliation{INFN, Sezione LNF, gruppo collegato di Cosenza}

\date{\today}

\begin{abstract}
Quantum communication protocols based on spin chains have been extensively studied, yet their ability to transmit nonstabilizer resources has not been systematically addressed. We investigate the transport of quantum magic in spin chains through the natural dynamics of systems initialized in nonstabilizer states, and quantify the transported resource via the stabilizer norm. We analyze three experimentally feasible state-transfer protocols, ranging from noisy to (quasi-)perfect transfer, including one realizable in trapped-ion platforms. We find that the geometry of the injected state strongly influences transport: states in the lower Bloch hemisphere achieve higher transfer quality, whereas states in the upper hemisphere give rise to an efficient magic transport only beyond a threshold value of the parameter controlling the tendency towards perfect transfer. These features are robust across all protocols and identify the Hamiltonian and state properties that favor high-quality transfer. Moreover, we identify a parameter region, relevant to the initial state preparation, in which the transported magic exceeds the initial encoding, indicating that such spin systems can act as magic-amplification channels. Our results establish the conditions for efficient transport of nonstabilizer resources and demonstrate quantum magic as a sensitive probe of quantum transport beyond population dynamics.

\end{abstract}

\maketitle

\textbf{\textit{Introduction}} --- Quantum state transfer (QST) is a well-established field of quantum information processing, motivated by the need to reliably transmit quantum information across scalable quantum architectures~\cite{bose2003quantum,christandl2004perfect,vinet2012how, apollaro2012fidelity, ahuja2026expedeting, christandl2017analytic, bezaz2025quasi, murphy2025ergotropy, dechiara2005from, chetcuti2020perturbative, michel2026quantum, campbell2011propagation}. Over the past two decades, extensive theoretical and experimental work has demonstrated that engineered spin chains can act as efficient quantum communication channels when the Hamiltonian parameters are appropriately tuned, enabling high fidelity transport of localized single excitation states as well as entanglement generation/distribution~\cite{chapman2016experimental, xiang2024enhanced, li2022quantum, zhou2024experimental}.

The performance of QST protocols is conventionally characterized by the average transfer fidelity, which quantifies how accurately an arbitrary input state is reconstructed at the receiving site. However, fidelity alone does not capture the transport of specific quantum resources. Of particular interest is nonstabilizerness (or magic), which is a fundamental resource for universal fault-tolerant quantum computation~\cite{bravyi2005universal,campbell2011catalysis,howard2017application,warmuz2025magic,haug2023stabilizer,veitch2014resource, sarkar2026all}. Motivated by its central role in quantum information processing, the generation, dynamics, and resilience of magic in many-body systems have recently attracted significant attention~\cite{leone2022stabilizer,Oliviero_2022,haug2023quantifying,tarabunga2023many,tirrito2024quantifying,warmuz2025magic,haug2026efficient,passarelli2024nonstabilizerness,haug2025probing,tirrito2025universal,tmisha2025robusteness,turkeshi2025magic,viscardi2026interplay,passarelli2025chaos,russomanno2025nonstabilizerness,scocco2026rise,odavic2023complexity,smith2025nonstabilizerness,sarkar2026reduced,zavatti2026quantum, martinezazcona2026magic}. Yet, despite the extensive use of spin chains as quantum communication channels and paradigmatic platforms for perfect state transfer (PST), their ability to transport magic remains unexplored. This raises a fundamental question: how is magic transmitted through quantum spin networks away from the ideal PST regime? In particular, are some magic states more robust under transport than others, and is the underlying mechanism governing magic transfer nontrivial? Here we address these questions by investigating magic transfer across a broad class of state-transfer protocols, ranging from ideal PST architectures to experimentally relevant imperfect-transfer regimes.

\begin{figure}[!t]
\centering
\includegraphics[width=\columnwidth]{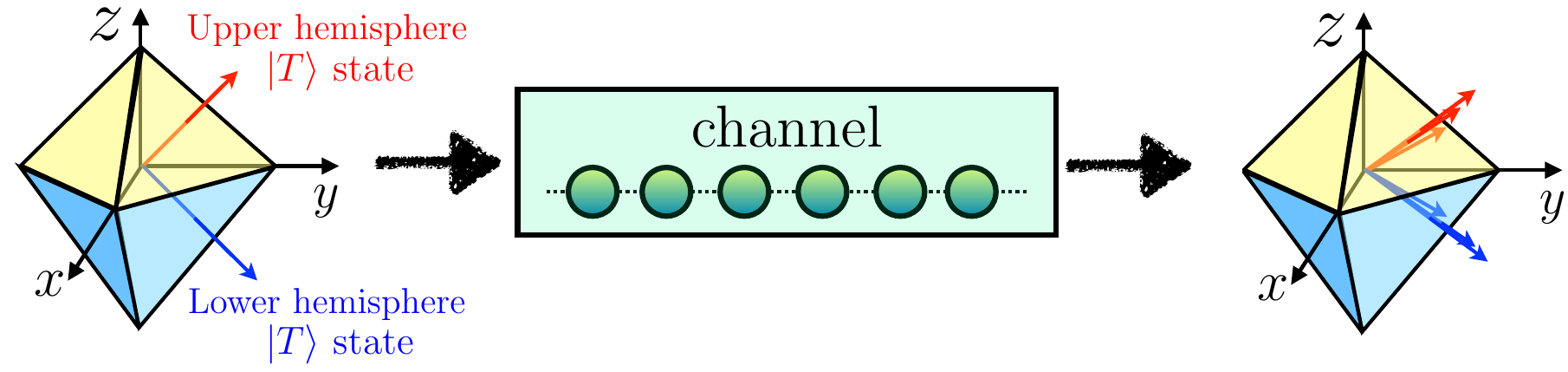}
\caption{Pictorial representation of the protocol: $\ket{T}$ states with maximum SN are injected at one edge of a quantum channel modeled as a spin chain and transferred to the opposite edge. The channel continuously interpolates between a perfect state transfer regime and an imperfect transfer regime. The transport of magic depends on the Bloch hemisphere to which the initial state belongs.}
\label{fig:sketch}
\end{figure}
\begin{figure*}[!t]
\centering
\includegraphics[width=\textwidth]{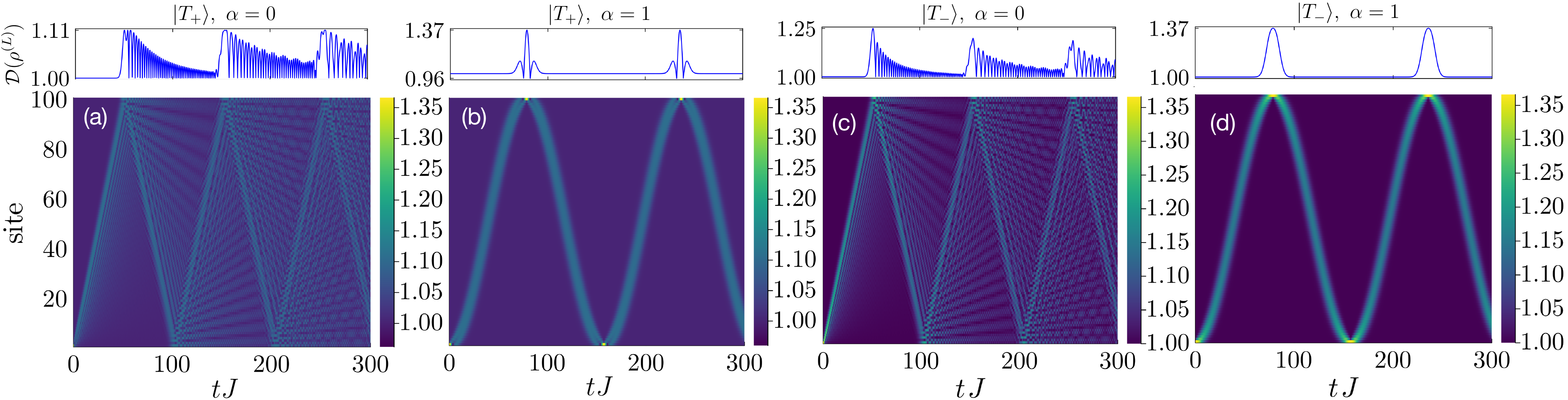}
\caption{On-site SN dynamics (colorbar) in a spin chain described by $\hat{H}_A$ composed of $L=100$ qubits in the two asymptotic regimes $\alpha=0$ and $\alpha=1$ when the system is initialized with $\ket{T_+}$ and $\ket{T_{-}}$ states in the site $j=1$. Specifically, panel (a) $\ket{T_+}, \, \alpha=0$, panel (b) $\ket{T_+}, \, \alpha=1$, panel (c) $\ket{T_-}, \, \alpha=0$, panel (d) $\ket{T_-}, \, \alpha=1$. On the top of each panel we report the SN dynamics of the $L$th site density matrix.}
\label{fig:dynamics}
\end{figure*}
We investigate the propagation of magic encoded in injected $T$-states, canonical single-qubit magic states enabling universal fault-tolerant quantum computation~\cite{bravyi2005universal}, through quantum spin chains acting as quantum buses, tuned away from perfect and quasiperfect state-transfer protocols. Considering three interpolation schemes connecting PST, quasiperfect state transfer (QPST), homogeneous, and long-range hopping models~\cite{christandl2004perfect,apollaro2012fidelity,franke2023quantum,Defenu_2023,fossfeig2025progress,browaeys2020many,barredo2015coherent}, we show that magic transport is governed by the geometry of quantum state space, exhibiting distinct regimes of suppression and robust propagation. Our findings further demonstrate that, depending on the initial state preparation and tuning parameters, the non-stabilizerness of the extracted state can exceed that of the initially encoded state. This reveals a novel class of spin chain quantum buses that, beyond faithfully routing information, intrinsically act as resource-amplification channels for quantum magic.

\textbf{\textit{Model and methods}} ---
We aim to investigate the transfer of nonstabilizerness across a quantum spin chain composed of $L$ qubits and described by a time-independent Hamiltonian $\hat{H}$ by analyzing its robustness under different state-transfer protocols, depending on whether the system operates close to known PST~\cite{christandl2004perfect}/QPST~\cite{apollaro2012fidelity} schemes or away from them. The protocol consists of injecting a non-stabilizer state $\ket{\psi(0)}$ at one edge of a quantum spin network, such that the initial global state is given by $\ket{\Psi(0)}=\ket{\psi(0)}\otimes\ket{\downarrow,\ldots,\downarrow}$, where all the remaining spins are initialized in the fully polarized state. The system then undergoes unitary evolution according to $\ket{\Psi(t)}=e^{-i\hat{H}t}\ket{\Psi(0)}$, with $\hbar=1$. To characterize the transport of magic through the chain, we analyze the nonstabilizerness of the reduced state at each site, $\rho^{(j)}(t)=\operatorname{Tr}_{\{1,\ldots,L\}\setminus j}\!\left[\ket{\Psi(t)}\bra{\Psi(t)}\right]$, with particular emphasis on the magic transferred to the opposite edge, described by the reduced state $\rho^{(L)}(t)$.

For single-qubit states, nonstabilizerness can be efficiently witnessed through the stabilizer norm (SN)~\cite{campbell2011catalysis, warmuz2025magic, haug2026efficient}, defined as $\mathcal{D}(\rho)=\frac{1}{2}\left(1 + |x| + |y| + |z|  \right)$, where $(x,y,z)$ are the components of the Bloch vector associated with the state $\rho$. Pure stabilizer states satisfy $\mathcal{D} ( \ket{\psi_{\rm st}}\bra{\psi_{\rm st}} )=1$, while all mixed stabilizer states obey $\mathcal{D} ( \rho_{\rm st} )\leq 1$, which we define as the polytope bound. Therefore, the condition $\mathcal{D} (\rho) > 1$ provides a necessary and sufficient criterion for nonstabilizerness. Geometrically, stabilizer states occupy a distinguished region of the Bloch sphere bounded by an octahedron whose six vertices correspond to the Pauli eigenstates located at $(\pm1,0,0)$, $(0,\pm1,0)$, and $(0,0,\pm1)$. Any state inside or on the surface of this octahedron is a stabilizer state, whereas states lying outside of it are nonstabilizer (magic) states. The SN is directly tied to the defining inequality of the octahedron, becoming larger than unity precisely when the state leaves the stabilizer region.

\begin{figure*}[!t]
\centering
\includegraphics[width=\textwidth]{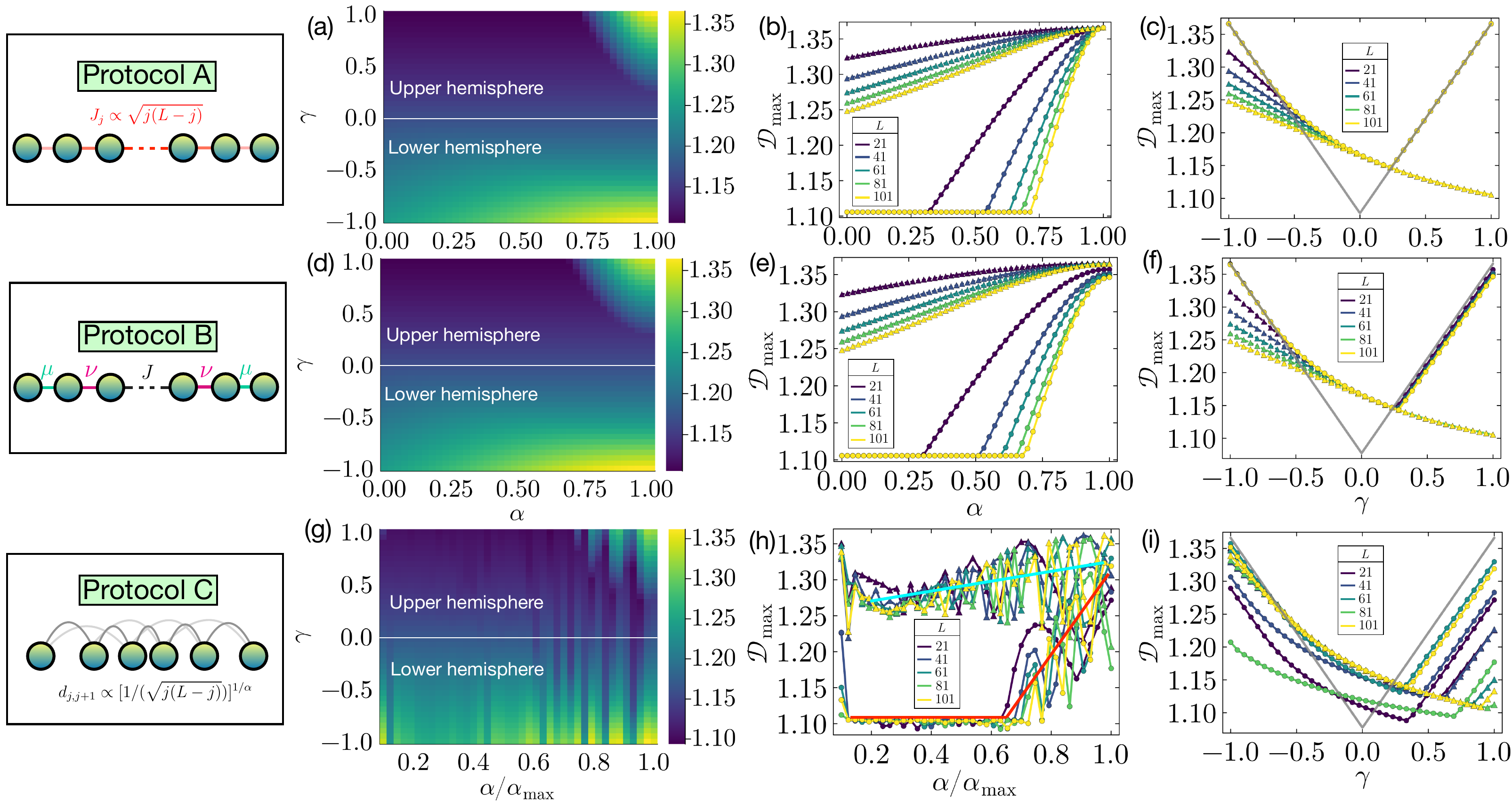}
\caption{\textit{Maximum transferable SN in the three protocols}. Panels (a)–(c) show results for protocol A (schematically illustrated on the left). Panel (a) reports the maximum transferable SN, $\mathcal{D}_{\rm max}$, as a function of the Hamiltonian parameter $\alpha$ and the initial-state parameter $\gamma$ for $L=101$. Panel (b) shows $\mathcal{D}_{\rm max}$ versus $\alpha$ for different system sizes $L$, comparing two initial states: $\gamma=-1$ ($\ket{T_-}$, triangles) and $\gamma=1$ ($\ket{T_+}$, circles). Panel (c) displays $\mathcal{D}_{\rm max}$ as a function of $\gamma$ for different values of $L$, with triangles denoting $\alpha=0$ and circles denoting $\alpha=1$ (PST). The gray curve indicates the SN of the first site initial state. In the PST case, for $\gamma<\gamma_*$, $\mathcal{D}_{\rm max}(\gamma)\simeq 0.074\, e^{-1.326\, \gamma} + 1.093$, while for $\gamma>\gamma_*$, $\mathcal{D}_{\rm max}(\gamma)\simeq \frac{1}{2}\left(1+\frac{2}{\sqrt{3}}\right)+\frac{\gamma}{2\sqrt{3}}$, with $\gamma_* \simeq 0.239$ estimated from the intersection of the two asymptotic expressions. Panels (d)-(f) and (g)–(i) show the analogous results for respectively protocol B and protocol C. In the latter case, the horizontal axis is rescaled as $\alpha/\alpha_{\rm max}$, with $\alpha_{\rm max}=5$ denoting the largest value considered; $\alpha/\alpha_{\rm max}=1$ corresponds to the point closest to the PST regime explored in our simulations. In panel (h), the cyan and red curves serve as guides to the eye, respectively, the $\gamma = -1$ and $\gamma = 1$ behaviors. In panel (i) triangles denote $\alpha=0.1$ and circles $\alpha=\alpha_{\rm max}=5$.}
\label{fig:maxstabbb}
\end{figure*}
We choose the state $\ket{\psi(0)}=\ket{T}$, where $\ket{T}$ denotes a pure single-qubit state whose Bloch vector has components $\frac{1}{\sqrt{3}}(\pm1,\pm1,\pm1)$, corresponding to maximal SN $\mathcal{D}(\ket{T} \bra{T}) = \frac{1+\sqrt{3}}{2}$. A representative choice is $\ket{T_+}=\cos\beta\ket{\uparrow}+e^{i\pi/4}\sin\beta\ket{\downarrow}$,
where the parameter $\beta$ is defined by $\cos(2\beta)=\frac{1}{\sqrt{3}}$. 
The Bloch vector of the initial state $\mathbf{r}^{(1)}_0=\Bigl[x^{(1)}_0,y^{(1)}_0,z^{(1)}_0\Bigr]$ is propagated through the chain dynamics to the Bloch vector $\mathbf{r}^{(j)}_t=\Bigl[x_t^{(j)},y_t^{(j)},z_t^{(j)}\Bigr]$ of the reduced density matrix at site $j$, according to the affine transformation
\begin{equation}
\mathbf{r}^{(j)}_t =\mathbf{M}^{(j)}_t\,\mathbf{r}^{(1)}_0 + \mathbf{c}^{(j)}_t,
\label{eq:affine}
\end{equation}
where $\mathbf{M}^{(j)}_t$ and $\mathbf{c}^{(j)}_t$ are respectively a matrix and a vector that depend on the single-excitation transition amplitude $f_j(t)=\braket{j|e^{-i\hat{H}t}|1}$  (see End Matter). Here,  $\ket{j}=\ket{\downarrow,\ldots,\uparrow_j,\ldots,\downarrow}$ denotes the single-excitation state localized at site $j$. This affine mapping directly connects the initial local Bloch vector to its time-evolved counterpart at any site $j$, thereby enabling a transparent description of the transport of local quantum features across the chain.
In particular, it allows for a direct evaluation of the on-site SN, $\mathcal{D} ( \rho^{(j)} (t) )=\frac{1}{2}\left(1 + |x_t^{(j)}| + |y_t^{(j)}| + |z_t^{(j)}| \right)$, which we use to track propagation of nonstabilizerness along the spin chain. Finally, we note that the time-evolved Bloch components can equivalently be obtained as expectation values of local Pauli operators at site $j$, i.e., $x_t^{(j)}=\langle \hat{\sigma}^x_j(t)\rangle$, $y_t^{(j)}=\langle \hat{\sigma}^y_j(t)\rangle$, and $z_t^{(j)}=\langle \hat{\sigma}^z_j(t)\rangle$, establishing a direct connection between the transfer matrix description and experimentally accessible observables. 

The magic transfer is studied within the zero- and single-excitation sectors of a quantum spin network governed by the XX Hamiltonian
\begin{equation}
\hat{H}=\sum_{i,j}J_{ij}
\left(
\hat{\sigma}_i^+\hat{\sigma}_j^-+\mathrm{H.c.}
\right),
\end{equation}
where \(\hat{\sigma}_j^\pm\) denote spin-raising and spin-lowering operators acting on site \(j\). We focus on one-dimensional spin chains and consider three state-transfer protocols that continuously connect imperfect to high-fidelity transfer.

\textit{Protocol A} interpolates between a homogeneous nearest-neighbor chain and the PST chain of Ref.~\cite{christandl2004perfect} through couplings \(J_j^{\rm NN}=(1-\alpha)J+\alpha\, a_j\), where \(\alpha\in[0,1]\) and \(a_j=(2J/L)\sqrt{j(L-j)}\). \textit{Protocol B} similarly interpolates between homogeneous and ballistic transfer~\cite{apollaro2012fidelity}, with nearest-neighbor couplings \(T_j=(1-\alpha)J+\alpha T_j^{(\mathrm{bal})}\). Here \(T_1^{(\mathrm{bal})}=T_{L-1}^{(\mathrm{bal})}=\mu\), \(T_2^{(\mathrm{bal})}=T_{L-2}^{(\mathrm{bal})}=\nu\), and \(T_j^{(\mathrm{bal})}=J\) otherwise, yielding near-unit transfer fidelity through engineering only the boundary couplings. \textit{Protocol C} is motivated by programmable quantum simulators based on trapped ions and Rydberg-atom arrays, where interactions naturally decay as a power law~\cite{franke2023quantum,Defenu_2023,fossfeig2025progress,browaeys2020many,barredo2015coherent, lewis2023ion}. We consider couplings \(J_{ij}=J/d_{ij}^{\alpha}\), where \(\alpha\) controls the interaction range. The qubit positions are chosen such that \(d_{j,j+1}=[2L/(\sqrt{j(L-j)})]^{1/\alpha}\), with \(d_{ij}=|x_i-x_j|\). In the limit \(\alpha\rightarrow\infty\), long-range couplings are suppressed and the resulting nearest-neighbor couplings approach the PST profile.

\textbf{\textit{Magic transfer}} ---
We find that the nonstabilizerness dynamics is determined by the hemisphere of the initial Bloch vector, $z=\pm 1/\sqrt{3}$. All four $T$-states within a given hemisphere exhibit identical SN dynamics (see End Matter). We therefore focus on two representative states, $\ket{T_+}$ and $\ket{T_-}$, with Bloch vectors $\frac{1}{\sqrt{3}}(1,1,1)$ and $\frac{1}{\sqrt{3}}(1,1,-1)$, respectively (Fig.~\ref{fig:sketch}). As shown below, the two hemispheres display markedly different transport behavior.

Fig.~\ref{fig:dynamics} shows the SN dynamics for systems evolving under the Protocol~A Hamiltonian starting from injected $T$ states in the upper ($\ket{T_+}$) and lower ($\ket{T_-}$) hemispheres of the Bloch sphere. 
For initial states in the upper hemisphere, $\mathcal{D}(\rho^{(j)}(t))$ remains comparatively small and can even temporarily drop below unity, signaling that the local Bloch vectors enter the stabilizer polytope during the evolution. Moreover, the relatively small magnitude of the SN indicates that the state at the last site hardly approaches a genuine $T$-like structure. Conversely, when the injected state lies in the lower hemisphere, the SN remains systematically larger, showing that the local Bloch vectors stay outside the stabilizer polytope and can approach states with a much stronger $T$-state character.

The insets in Fig.~\ref{fig:dynamics} (top panels) show the time evolution of $\mathcal{D}(\rho^{(L)}(t))$, highlighting clear qualitative differences between the four cases. We first consider the $\alpha=0$ regime [Fig.~\ref{fig:dynamics}(a),(c)], where reflection times correspond to boundary scattering events of the propagating wave packets. For the initial state $\ket{T_+}$, the SN exhibits only modest growth, reaching $\mathcal{D}(\rho^{(L)})\approx 1.11$ at the first reflection and remaining within a narrow quasiperiodic window thereafter. In contrast, for $\ket{T_-}$ it reaches a significantly larger value, $\mathcal{D}(\rho^{(L)})\approx 1.25$, already at the first reflection, before decreasing at later times. In the PST limit $\alpha=1$ [Fig.~\ref{fig:dynamics}(b),(d)], $\mathcal{D}(\rho^{(L)}(t))$ becomes strictly periodic and repeatedly attains its maximal value, reflecting perfect revivals. Nevertheless, the two initial states display distinct trajectories: for $\ket{T_-}$, nonstabilizerness is transported to the opposite edge without crossing the stabilizer boundary, while for $\ket{T_+}$ the evolution enters the stabilizer region, as indicated by $\mathcal{D}(\rho^{(L)}(t))<1$, before reaching its maximum.

This behavior admits a clear geometric interpretation on the Bloch sphere. The last site is initialized in the $\ket{\downarrow}$ state in the lower hemisphere and evolved toward a target $\ket{T_\pm}$ state. For target states in the lower hemisphere, the unitary dynamics connect $\ket{\downarrow}$ to the target along shorter effective trajectories that remain in regions of high nonstabilizerness, enabling efficient magic transport. By contrast, for upper-hemisphere target states the evolution follows longer paths that approach the stabilizer region, where it becomes effectively confined to low-magic sectors, thereby suppressing the transfer of nonstabilizerness.

\textbf{\textit{Maximum transferable SN}} --- Here, we quantify the maximum SN transferred between the chain edges. Since SN witnesses nonstabilizerness through violation of the stabilizer-polytope bound, larger transferred values indicate better preservation of magic during transport and stronger retention of $T$-state-like features.

We quantify the maximum transferable SN  $\mathcal{D}_{\rm max}=\max_{t\in [0,T]} \mathcal{D}(\rho^{(L)}(t))$ over a time window $T$ chosen to be sufficiently long to capture multiple reflection events of the excitation. To estimate the relevant dynamical timescale, we use the PST reflection time $\tau_{\rm ref} J = \frac{\pi L}{4}$~\cite{christandl2004perfect} and set the total observation window to $T = N \tau_{\rm ref}$, with $N=10$ chosen to ensure that several reflection periods are included in the analysis.
For a fixed protocol, $\rho^{({L})}(t)$ depends on the parameter $\alpha$ and on the choice of the single-qubit state encoded at the initial time $t=0$.
The previous section examined magic transfer for the two states $\ket{T_+}$ and $\ket{T_-}$. Here, we extend the analysis by studying the maximum transferable stabilizer norm for input states that are, in general, mixed, of the form  $\rho^{(1)}_{\gamma}(0)= \frac{1}{2}\left(\hat{\mathbb{I}} + \frac{1}{\sqrt{3}}\hat{\sigma}^x_1 + \frac{1}{\sqrt{3}}\hat{\sigma}^y_1 + \frac{\gamma}{\sqrt{3}} \hat{\sigma}_1^z\right)$, with $\gamma \in [-1,1]$ interpolating between $\ket{T_-}$ and $\ket{T_+}$. Thus, $\mathcal{D}_{\rm max}=\mathcal{D}_{\rm max}(\alpha, \gamma)$, exhibiting nontrivial features in the parameter space $(\alpha,\gamma)$.

Our results are summarized in Fig.~\ref{fig:maxstabbb}, for the three aforementioned protocols (see End Matter for the specific values of $\mu$ and $\nu$ for Protocol B). We first observe that the maximum transferable SN depends nontrivially on both the initial state preparation ($\gamma$) and the Hamiltonian configuration ($\alpha$) across all three protocols [Figs.~\ref{fig:maxstabbb}(a),(d),(g)]. States initialized in the upper hemisphere exhibit a clear crossover behavior characterized by a threshold $\alpha_*$ that must be exceeded to achieve high-quality magic transfer. Conversely, for initial states in the lower hemisphere, $\mathcal{D}_{\rm max}$ displays smooth behavior. Physically, this means that in the first case the system starts in a down-spin state at the last site, so a threshold must be overcome before the coupling to neighboring spins displaces the Bloch vector sufficiently far from the stabilizer polytope, enabling efficient transfer of magic. In addition, we observe a common behavior across all three protocols, with protocol C exhibiting more erratic behavior, consistent with prior work on the effects of long-range interactions on perfect state transfer \cite{Ronke_2011}. As the ballistic scheme (protocol B) provides speed advantages over the transfer times required for PST~\cite{Yung_2006,apollaro2012fidelity}, this appears to be indicative of the existence of an explicit class of quantum states capable of propagating through the system more rapidly without being taxed by as substantial a reduction in the faithfulness of the transport.

The presence of the threshold $\alpha_*$ for initial $\ket{T_+}$ states can be better appreciated in Figs.~\ref{fig:maxstabbb}(b),(e),(h), where $\mathcal{D}_{\rm max}(\alpha)$ is reported for different system sizes $L$ for both $\ket{T_+}$ and $\ket{T_-}$. We observe that protocols A and B share approximately the same threshold location, with the latter increasing with $L$, suggesting high-quality transfer only in the $\alpha \to 1$ limit as $L \to \infty$. Protocol C instead exhibits a more erratic behavior, with no clear size dependence of the threshold. In any case, we find $\alpha_* \gtrsim 3$, suggesting that relevant magic transfer emerges only when long-range effects become sufficiently suppressed.

Finally, in Fig.~\ref{fig:maxstabbb}(c),(f),(i), we report $\mathcal{D}_{\rm max}(\gamma)$ for different system sizes at fixed $\alpha=0$ and $\alpha=1$ (PST) for protocols A and B, and at $\alpha=0.1$ and $\alpha=5$ for protocol C. Tuning $\gamma\in[-1,1]$ corresponds to moving from the lower to the upper hemisphere. The quantity $\mathcal{D}_{\rm max}(\gamma)$ is compared with $\mathcal{D}(\rho^{(1)}(0))$, shown in gray. For protocols A and B, we observe a clear distinction between the noisy and PST (or QST) regimes. In the former, $\mathcal{D}_{\rm max}$ decreases monotonically with $\gamma$ and shows a strong system-size dependence in the lower hemisphere, confirming that the upper hemisphere corresponds to low-quality magic transfer. In contrast, the PST (and QPST) regime behaves markedly differently: $\mathcal{D}_{\rm max}(\gamma)$ is well described by an exponentially decaying
branch separated by a crossover point $\gamma_*\simeq 0.239$ from a linearly increasing branch. In this regime, $\mathcal{D}_{\rm max}$ follows the same behavior as the initial state, indicating that the maximum transferable stabilizer norm coincides with that of the injected state. Here, the dependence on system size is absent or negligible. We also identify a region of $\gamma$ in the central part of the Bloch sphere where the maximum transferable SN exceeds that of the initial state, indicating a form of magic enhancement associated with the state being driven deep into the non-stabilizer region, far from the stabilizer polytope. For protocol C, we do not observe a clear $L$ dependence. The separation between the monotonically decreasing and linearly increasing regimes, marked by a crossover point, persists for both values of $\alpha$ considered.

\textbf{\textit{Conclusions and outlook}}---
In this work, we investigated the transport of quantum magic through spin chain quantum buses, quantifying the transferred resource via the stabilizer norm. To probe genuinely nontrivial transport regimes, we focused on buses whose dynamics do not, in general, coincide with those supporting  PST or QPST, except for specific parameter choices. In particular, we analyzed two protocols, A and B, controlled by a parameter $\alpha$ that interpolates between homogeneous couplings and PST/QPST-supporting configurations, together with a third protocol, C, motivated by experimentally realizable trapped-ion architectures.

We find that magic transport is strongly influenced by the geometry of the initial state on the Bloch sphere. While magic encoded in states belonging to the lower hemisphere is efficiently transferred across a broad parameter range, states in the upper hemisphere exhibit a transport threshold in $\alpha$, below which magic transfer is strongly suppressed. This reveals a marked asymmetry in the transport properties of quantum magic that is absent in conventional state-transfer figures of merit.
This asymmetry is particularly evident for $T$-states. Upper-hemisphere $T$-states exhibit transport thresholds that increase with the size of the system, requiring progressively finer tuning and eventually approaching regimes compatible with PST/QPST. In contrast, lower-hemisphere $T$-states do not show threshold behavior and present larger maximum transferable stabilizer norm.

Furthermore, for nonmaximally magic input states, the conditions maximizing state-transfer fidelity do not generally coincide with those maximizing the transported magic. More strikingly, we identify a parameter regime in which the quantum bus not only transfers but also amplifies the injected magic, yielding a stabilizer norm at the receiving end larger than that of the initially encoded state.
These results establish quantum-magic transport as a distinct resource-transfer problem beyond conventional PST-based paradigms and demonstrate that experimentally relevant spin-chain architectures can exhibit nontrivial resource-processing capabilities.

Future work may address the effects of disorder, finite temperature, dissipation, and repeated bus operation on the transport and enhancement of quantum magic. Understanding these effects is also essential from an experimental perspective, where the transfer of non-stabilizer resources may be realized in superconducting circuits~\cite{zhou2024experimental, li2018perfect}, photonic platforms~\cite{chapman2016experimental}, trapped-ion systems~\cite{fossfeig2025progress, lewis2023ion}, and atomtronic networks~\cite{polo2024perspective}, enabling the distribution of quantum computational resources across quantum spin networks.

\textbf{\textit{Acknowledgements}} --- AP and FP acknowledge O. La Rocca  for valuable discussions during the development of this work. AP and FP acknowledge support from PNRR MUR Project No. PE0000023-NQSTI through the secondary project ThAnQ. CN acknowledges support from EPSRC, grant number is EP\textbackslash W524657\textbackslash1. ECD acknowledges support from the European
Unions Horizon Europe research and innovation programme
under grant agreement No 101114305 (MILLENION-
SGA1 EU Project), from PID2021-127726NB-I00 and
PID2024-161474NB-I00 (MCIU/AEI/FEDER,UE), and from
QUITEMAD-CM TEC-2024/COM-84. ECD also acknowledges support from the Grant IFT Centro de Excelencia Severo Ochoa CEX2020-001007-S funded by
MCIN/AEI/10.13039/501100011033, and from the CSIC Research Platform on Quantum Technologies PTI-001.

\section*{End Matter}
\label{app:bloch_vec}

\textbf{\textit{Time evolution of local Bloch vectors}}--- In the following we briefly review the derivation of Eq.~\ref{eq:affine}, which describes the time evolution of the Bloch vector. The initial single-qubit state is fully specified by the Bloch vector components $\mathbf{r}_0^{(1)}=[x_0^{(1)},y_0^{(1)},z_0^{(1)}]$, which determine all matrix elements of the density operator in the computational basis. In particular, the diagonal elements (populations) are given by
\begin{equation}
\rho_{\uparrow\uparrow}^{(1)}(0)=\frac{1+z_0^{(1)}}{2},
\qquad
\rho_{\downarrow\downarrow}^{(1)}(0)=\frac{1-z_0^{(1)}}{2},
\end{equation}
while the off-diagonal coherences are
\begin{equation}
\rho_{\uparrow\downarrow}^{(1)}(0)=\frac{x_0^{(1)}-i y_0^{(1)}}{2},
\qquad
\rho_{\downarrow\uparrow}^{(1)}(0)=\frac{x_0^{(1)}+i y_0^{(1)}}{2}.
\end{equation}

Starting from these initial conditions, the reduced density matrix at site $j$ and time $t$ under the XX spin-chain dynamics \cite{campbell2011propagation,murphy2025ergotropy} takes the form
\begin{equation}
\rho^{(j)}(t)=
\begin{pmatrix}
\rho_{\uparrow\uparrow}^{(1)}(0)|f_j(t)|^2
&
\rho_{\uparrow\downarrow}^{(1)}(0)f_j(t)
\\
\rho_{\downarrow\uparrow}^{(1)}(0)f_j^*(t)
&
\rho_{\downarrow\downarrow}^{(1)}(0)
+\rho_{\uparrow\uparrow}^{(1)}(0)\left(1-|f_j(t)|^2\right)
\end{pmatrix},
\end{equation}
where $f_j(t)=f_j^R(t)+i f_j^I(t)=\braket{j|e^{-i\hat{H}t}|1}$ is the single-excitation transition amplitude associated with the Hamiltonian $\hat{H}$. This allows the evolved state to be expressed in Bloch form as 
\begin{equation} 
\rho^{(j)}(t)=\frac{1}{2}\left(\hat{\mathbb{I}}+x_t^{(j)}\hat{\sigma}_j^x+y_t^{(j)}\hat{\sigma}_j^y+z_t^{(j)}\hat{\sigma}_j^z\right), 
\end{equation} 
where the time-dependent Bloch components are defined as $a_t^{(j)}=\braket{\hat{\sigma}_j^a(t)}$, with $a=x,y,z$. By comparing the matrix elements, one finds that the transverse components evolve according to 
\begin{equation} 
\begin{aligned} 
x_t^{(j)} &= x_0^{(1)} f_j^R(t)+y_0^{(1)} f_j^I(t),\\ y_t^{(j)} &= -x_0^{(1)} f_j^I(t)+y_0^{(1)} f_j^R(t), \end{aligned} 
\end{equation} 
while the longitudinal component evolves independently as 
\begin{equation} 
z_t^{(j)}=(1+z_0^{(1)})|f_j(t)|^2-1. 
\end{equation} 
In compact form, the equation of motion can be written as 
\begin{equation} 
\mathbf{r}^{(j)}_t=\mathbf{M}^{(j)}_t\,\mathbf{r}^{(1)}_0+\mathbf{c}^{(j)}_t, \label{eq:Bloch_vector} 
\end{equation} 
where 
\begin{align} \mathbf{M}^{(j)}_t &= \begin{bmatrix} f_j^R(t) & f_j^I(t) & 0 \\ -f_j^I(t) & f_j^R(t) & 0 \\ 0 & 0 & |f_j(t)|^2 \end{bmatrix}, \\ \mathbf{c}^{(j)}_t &= \begin{bmatrix} 0 \\ 0 \\ |f_j(t)|^2 - 1 \end{bmatrix}, 
\end{align} 
and $\mathbf{r}^{(1)}_0=[x^{(1)}_0,y^{(1)}_0,z^{(1)}_0]$ and $\mathbf{r}^{(j)}_t=[x_t^{(j)},y_t^{(j)},z_t^{(j)}]$ denote the initial Bloch vector at the first site and the time-evolved Bloch vector at site $j$, respectively.

We now consider the $j$th site SN
\begin{equation}
\mathcal{D}(\rho^{(j)}(t))
=
\frac{1}{2}\left(
1 + |x_t^{(j)}| + |y_t^{(j)}| + |z_t^{(j)}|
\right),
\end{equation}
and focus on initial states with
\begin{equation}
\mathbf{r}^{(1)}_0=\frac{1}{\sqrt{3}}(\pm 1,\pm 1,\pm 1),
\end{equation}
namely the $T$-states, which maximize the SN. We observe that, once $z_0^{(1)}$ is fixed, thereby selecting the hemisphere to which the initial state belongs, the SN is invariant under the remaining sign choices. Therefore, to determine its time evolution, it is sufficient to analyze only one of the four states within a given hemisphere.

\begin{figure}[h!]
    \centering
    \includegraphics[width=0.5\linewidth]{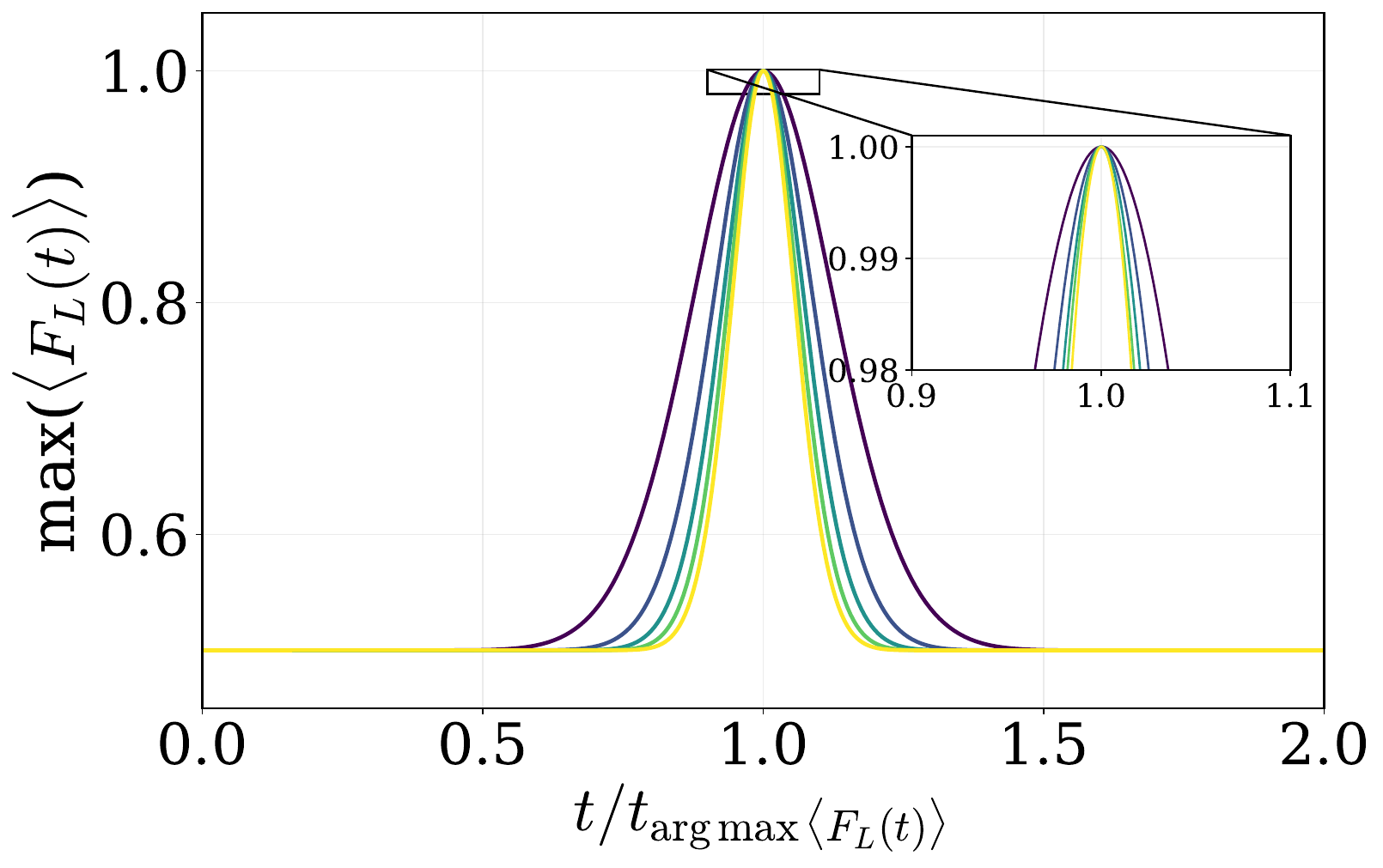}\includegraphics[width=0.50\linewidth]{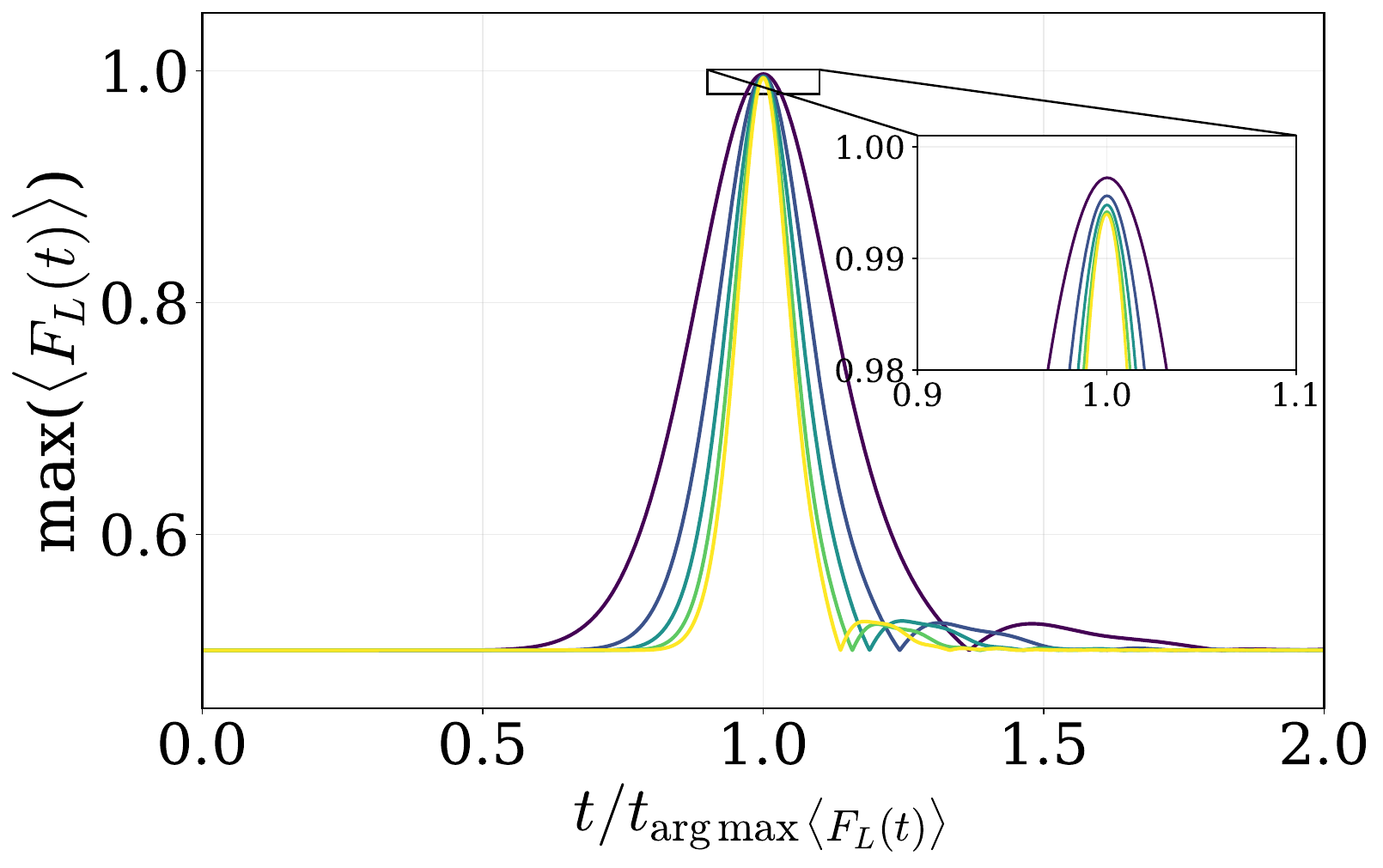}
    \includegraphics[width=0.51\linewidth]{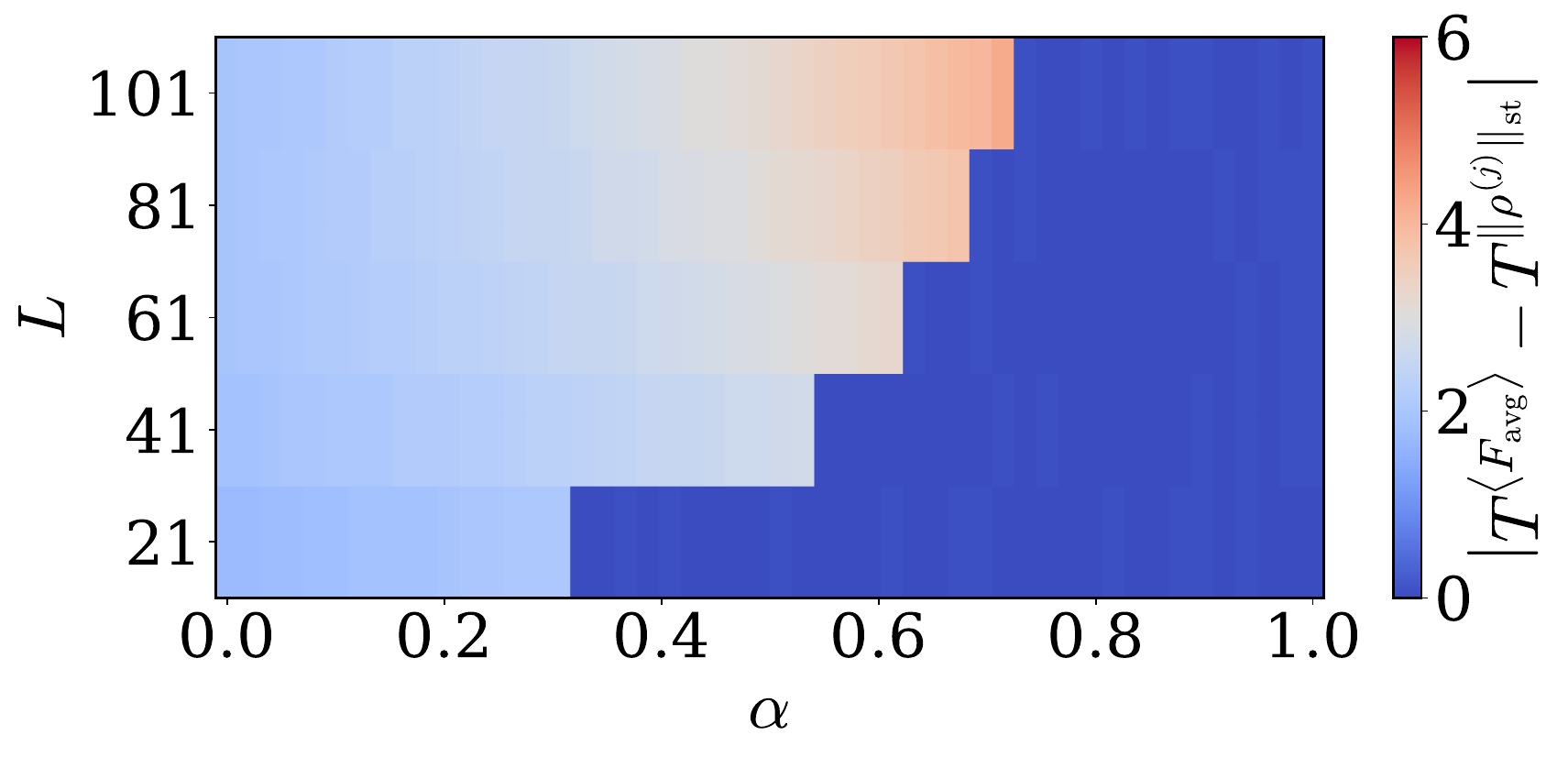}\includegraphics[width=0.51\linewidth]{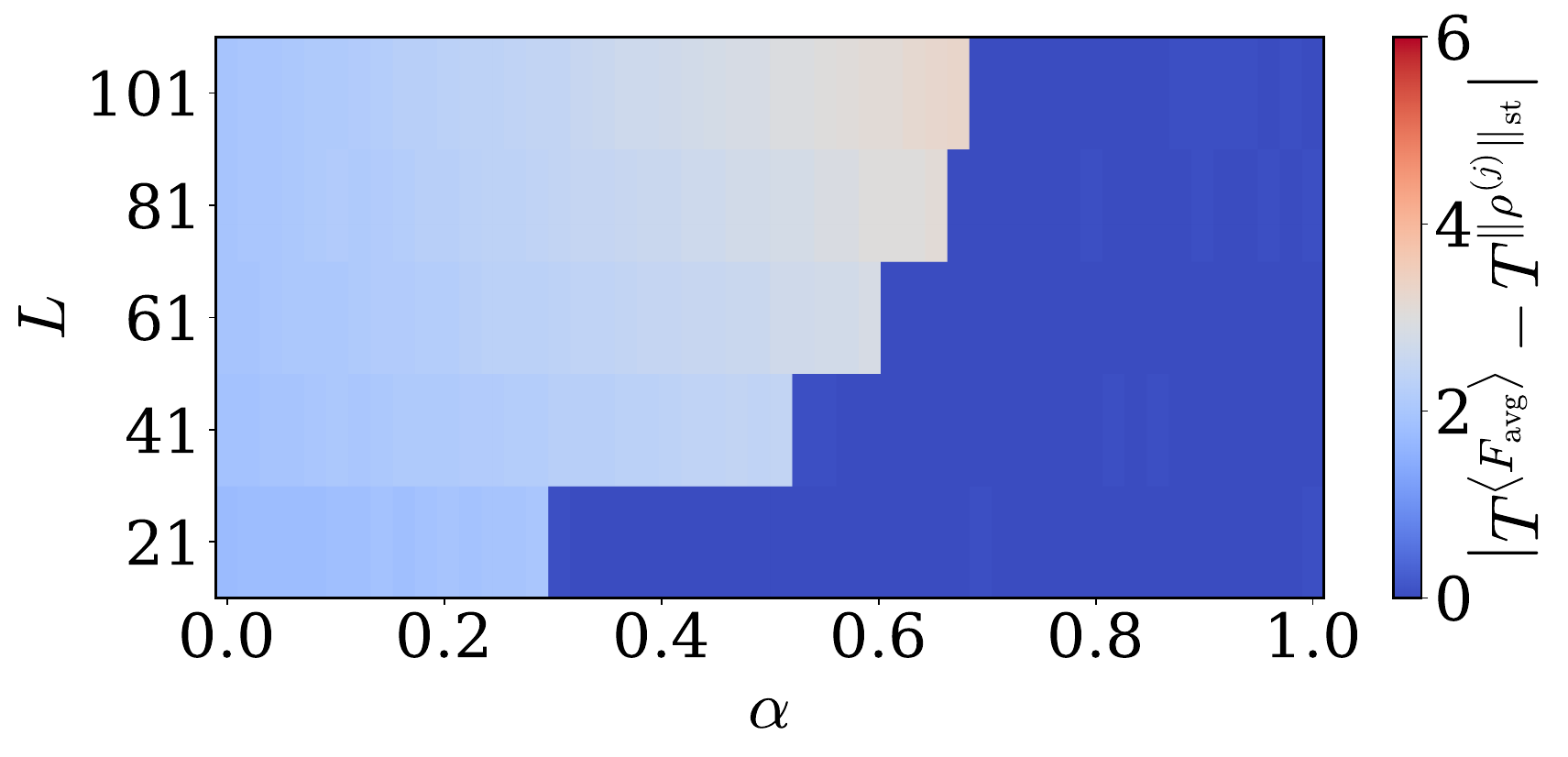}
    \caption{Maximal average fidelity over normalised time units ($J^{-1}$) where the maximal fidelity occurs for $\alpha=1$ (top row). These varying $L$-solutions correspond to \textit{Protocol A} (left column) and \textit{Protocol B} (right column). Bottom row corresponds to the difference in times between maximum SN and the maximal average fidelity on the $L^{th}$ site.}
    \label{fig:4}
\end{figure}

\textbf{\textit{Average fidelity and chain parametrization}}---The canonical figure-of-merit characterizing the overall quality of the chain as a quantum communication channel (independent of stabilizerness) is the so-called ``average fidelity" 
\begin{equation}
\braket{F_j(t)}=\frac{1}{2}+\frac{\left|f_j(t)\right|\text{cos}(\gamma_j)}{3}+\frac{\left|f_j(t)\right|^2}{6}, 
\label{avgfid}
\end{equation}
where $\gamma_j = \text{arg}(f_j(t))$ denotes the phase of the transition amplitude, which is obtained by averaging the fidelity over all pure input states in the Bloch sphere \cite{bose2003quantum,NikolopoulosJex_2014}. We note that $\gamma_j$ in Eq.~\ref{avgfid} may be adjusted such that it is a multiple of $2\pi$, therein maximizing $\braket{F_j(t)}$. This can be done by modifying the magnetic field profile of the system and/or applying an appropriate rotation (such as a phase gate $R_z(\theta)$) on the receiver qubit.
Fig.~\ref{fig:4} shows the peak average fidelity for Protocols A and B, with time normalized as $t/t_{\arg\max \langle F_L(t)\rangle}$, such that the fidelity peak occurs at $t=1$ for all chain lengths. 

Fig.~\ref{fig:maxstabbb} in the main text shows that the time at which the average fidelity reaches its maximum generally does not coincide with the time at which the stabilizer norm is maximized. This mismatch can arise either because the quantum bus operates away from the PST (QPST) regime or because the initially encoded nonstabilizerness is nonmaximal, even when the transfer channel itself is close to ideal tuning [see Fig.~\ref{fig:maxstabbb}(c), (f), and (i)].

The bottom panel of Fig.~\ref{fig:4} shows the time difference between the peak average fidelity, occurring at $T^{\langle F_{\mathrm{avg}} \rangle}$, and the maximum stabilizer norm on the final site, occurring at $T^{\langle ||\rho^L||_{\mathrm{st}} \rangle}$; maximum SN time and average fidelity time coincide at $\alpha > \alpha_*$. However, below this threshold, the time that maximizes SN does not correspond to the time at which the state is, on average, best transferred.

\textbf{\textit{Protocol B optimized parameters}}---
\begin{table}[h!]
\centering
\caption{\label{app:tab} Values of $\mu$ and $\nu$ for different system sizes $L$.}
\begin{tabular}{|c|c|c|}
\toprule
$L$ & $\mu$ & $\nu$ \\
\midrule
21  & 0.53833 & 0.80753 \\
\hline
41  & 0.45935 & 0.75485 \\
\hline
61  & 0.40854 & 0.71366 \\
\hline
81  & 0.38936 & 0.70472 \\
\hline
101 & 0.35840 & 0.67420 \\
\bottomrule
\end{tabular}
\end{table}
Table.~\ref{app:tab} shows the optimized values of $\mu$ and $\nu$ for QPST (ballistic transport) under Protocol B across the chain lengths considered. These values were not originally reported in Ref.~\cite{apollaro2012fidelity}; instead, they were obtained by optimizing the outer two couplings in otherwise homogeneous spin chains to maximize the transfer fidelity within a target time window of $t \cdot J = \pi\sqrt{L^2-1}/4$ (odd chain speed limit). The optimization was performed using \texttt{scipy.optimize.differential\_evolution} in SciPy~\cite{Virtanen2020}.
\end{document}